\documentclass{cdclass}

\usepackage[bibstyle=numeric,
			citestyle=numeric-comp,
			natbib=true,
			backend=biber,
			maxcitenames=12,
		    mincitenames=10,
			sorting=none,]{biblatex}

\begin{document}

\title{Together apart: the influence of increased crowd heterogeneity on crowd dynamics at bottlenecks}
\titlerunning{Influence of increased crowd heterogeneity on crowd dynamics at bottlenecks} 

\author{
  Paul Geoerg\authorlabel{1} \and 
  Ann Katrin Boomers\authorlabel{2} \and 
  Maxine Berthiaume\authorlabel{3,4} \and 
  Maik Boltes\authorlabel{2} \and 
  Max Kinateder\authorlabel{4,5} 
}
\authorrunning{P. Geoerg \and A. Boomers \and M. Berthiaume \and M. Boltes \and M. Kinateder}
\institute{
  \authorlabel{1} Vereinigung zur Förderung des Deutschen Brandschutzes e.V., Münster, Germany,
  \authoremail{1}{geoerg@vfdb.de}
  \and
  \authorlabel{2} Forschungszentrum Jülich, Jülich, Germany
  \authoremail{2}{m.boltes@fz-juelich.de, a.boomers@fz-juelich.de}
  \and
  \authorlabel{3} University of Ottawa, Ottawa, Canada 
    \authoremail{3}{mbert094@uottawa.ca}
    \and
  \authorlabel{4} National Research Council Canada, Ottawa, Canada, 
    \authoremail{4}{max.kinateder@nrc-cnrc.gc.ca}
    \and
  \authorlabel{5} Carleton University, Ottawa, Canada   
}

\date{year}{date1}{date2}{date3} 
\ldoi{10.17815/CD.20XX.X                                                                                                          } 
\volume{V}  
\online{AX} 

\maketitle

\begin{abstract}
Individual differences in mobility (e.g., due to wheelchair use) are often ignored in the prediction of crowd movement. Consequently, engineering tools cannot fully describe the impact of vulnerable populations on egress performance. To contribute to closing this gap, we performed laboratory experiments with \SI{25}{} pedestrians with varying mobility profiles. The \textit{control condition} comprised only participants without any additional equipment; in the \textit{luggage condition} and the \textit{wheelchair condition}, two participants at the center of the group either carried suitcases or used a wheelchair. We found that individuals using wheelchairs and to a lesser degree those carrying luggage needed longer to pass through the bottleneck, which also affected those walking behind them. This led to slower times to fully clear the bottleneck in the wheelchair and luggage condition compared to the control group. The results challenge the status quo in existing approaches to calculating egress performance and other key performance metrics in crowd dynamics. 
\end{abstract}


\clearpage
\newpage

\section{Introduction}
\label{sec:introduction}

Crowd accidents are a frequent occurrence on a global scale and can have severe consequences for pedestrian health and safety, and numerous studies have been published on crowd movement and safety \cite{Feliciani.2023, Haghani.2023b}. As a result, the quality and quantity of knowledge in this area have significantly increased in recent years \cite{Gwynne.2016, Haghani.2020c, Haghani.2021}. Nevertheless, the variability in mobility among individuals has frequently been overlooked. (e.g., \cite{Wang.2021}). This is critical because people with reduced mobility (e.g., due to temporary or permanent disabilities) represent a large part of our societies, and variability in mobility likely influences pedestrian dynamics. In 2011, approximately \SI{15.6}{\percent} of the world's population were living with disabilities \cite{WorldHealthOrganization.2011}. Of those, about \SI{10}{\percent} used a wheelchair as a mobility device \cite{WorldHealthOrganization.2010}. These statistics likely have changed, given the global trend of aging societies and the correlation between increasing age and risk of living with a disability \cite{Tas.2007}. 

These trends have consequences for performance-based analysis of pedestrian movement. The prediction and evaluation of egress performance in the built environment is a crucial element of performance-based design in fire safety engineering. However, estimated capacities and key performance values usually rely on data from homogeneous crowds, i.e., data from young adults without disabilities. Individual differences in mobility, space requirements (e.g., due to wheelchair use), and interactions between people are ignored. Therefore, engineering tools lack ecological validity and cannot adequately describe egress performance. Several studies observed reduced flows through bottlenecks and lower pedestrian densities in heterogeneous crowds \cite{Shimada.2006, Sharifi.2014, Sharifi.2015b, Sharifi.2016, Sharifi.2017, Pan.2020, Pan.2021}. A review of engineering egress data considering people with disabilities suggested supplementing pre-movement and horizontal movement data sets considering different types of mobility characteristics \cite{Geoerg.2019d, Gwynne.2016}. 

Several studies reported reduced movement speed through bottlenecks in crowds with wheelchair users compared to data from more homogeneous samples \cite{Tsuchiya.2007, Fu.2022b, Geoerg.2019, Geoerg.2019b, Tanimoto.2011}. Tsuchiya et al. found that if the bottleneck was wide enough to allow pedestrians to pass, wheelchair users were overtaken and the general speed increased. However, when wheelchair users passed the bottleneck, both the local density and the flow were reduced. The observed effect scaled with the number of wheelchair users and reduced the flow by \SI{15}{\percent} (with up to three wheelchair users present) \cite{Tsuchiya.2007}. 
Similar trends were reported by Daamen et al., who observed a decreased capacity of \SI{20}{\percent} of the bottleneck when three participants with simulated blindness and three participants in wheelchairs were part of the study population \cite{Daamen.2010}. 

In one study, wheelchair users made up \SI{13}{\percent} of the sample and led to a reduced overall movement speed \cite{Tanimoto.2011}. This study also compared space requirements between participants in wheelchairs and physical dimensions of wheelchairs: While the physical area of wheelchairs used in the study was reported as approx \SI{0.3}{\square\meter}, the approximated required space for free movement during trials was reported as about \SI{0.8}{\square\meter}. These findings are in line with previous work that reported larger spatial distribution areas of wheelchairs than of pedestrians \cite{Yayun.2023}. It should be noted, however, that wheelchairs and other assistive devices can vary significantly in appearance, dimensions, and how they are operated.

The movement characteristics of wheelchair users differ from pedestrians and also affect movement in their neighborhood. Overall, Feliciani and colleagues \cite{Feliciani.2020} observed smoother movement patterns for wheelchair users and their neighbors. However, Shimada et al. only observed a smoother movement of self-operated wheelchair users and not in assisted operated wheelchairs \cite{Shimada.2006}.

An important question is what the underlying mechanisms of the differences between heterogeneous and homogeneous crowds are. Previous work has found anecdotal evidence that pedestrians keep a larger distance from wheelchair users \cite{Geoerg.2022}. This change in behavior may have cascading effects on microscopic and macroscopic movement parameters. One potential explanation is that pedestrians without disabilities attribute higher \textit{vulnerability} to people with visible disabilities, and therefore are more courteous when moving near them. However, to the best of our knowledge, the effects of perceived vulnerability as well as the required space of others have not yet been studied in the context of crowd movement.


To address this gap, we present results from controlled experiments studying the movement of crowds with a fixed number of wheelchair users (higher perceived vulnerability and space requirements), participants with luggage (higher space requirements), or without any devices moving through a bottleneck.

\section{Materials and methods}
\label{sec:methods}

\subsection{Pilot study}
A pilot study was conducted to assess the perceived vulnerability of a range of pedestrian profiles (see \cite{Geoerg.2023g}). \SI{51}{} participants compared persons with different mobility attributes in a Two-Alternatives-Forced-Choice (2AFC) online survey. Participants were shown pairs of images, each showing a illustrated character, and then asked to select the person who appeared more vulnerable to them. For example, participants would choose between a male person in a wheelchair and a female person carrying a suitcase. In total \SI{16}{} different stimuli (male vs. female; no device, 1 suitcase, 2 suitcases, small backpack, large backpack, stroller, cane, and wheelchair), yielding $n(n-1)/2 = 120$  pairwise comparisons per participant. The detailed methods and procedures of the pilot study are described \cite{Geoerg.2023g}. \fref{fig:perceivedvulnerability} shows that wheelchair users appeared the most vulnerable and persons without any items/devices the least vulnerable. Persons carrying two suitcases were in the middle. These results informed the design of the main study, in which we selected these three conditions.

\subsection{Main study}
Data from up to \SI{25}{} volunteers were collected on three consecutive days in an indoor gym. All participants wore hats with fixed ArUco Markers which provide a unique identifier, positional and directional information. Participants were recorded with calibrated overhead cameras with a framerate of \SI{30}{\per\second}. The head trajectories were extracted from the raw video footage using PeTrack \cite{ Boltes.2022}.

Participants were primarily young adults between 21 and 45 years (mean age: \SI[parse-numbers = false]{28.5 \pm 11.7}{} years). \SI{54}{\percent} identified as female, \SI{45}{\percent} as male, and \SI{1}{\percent} with another gender identity. All participants gave informed consent and were compensated for their time. The study was approved by the NRC Research Ethics Board (REB2021-106).

\begin{figure}
  \centering
  \includegraphics[width=1.0\linewidth]{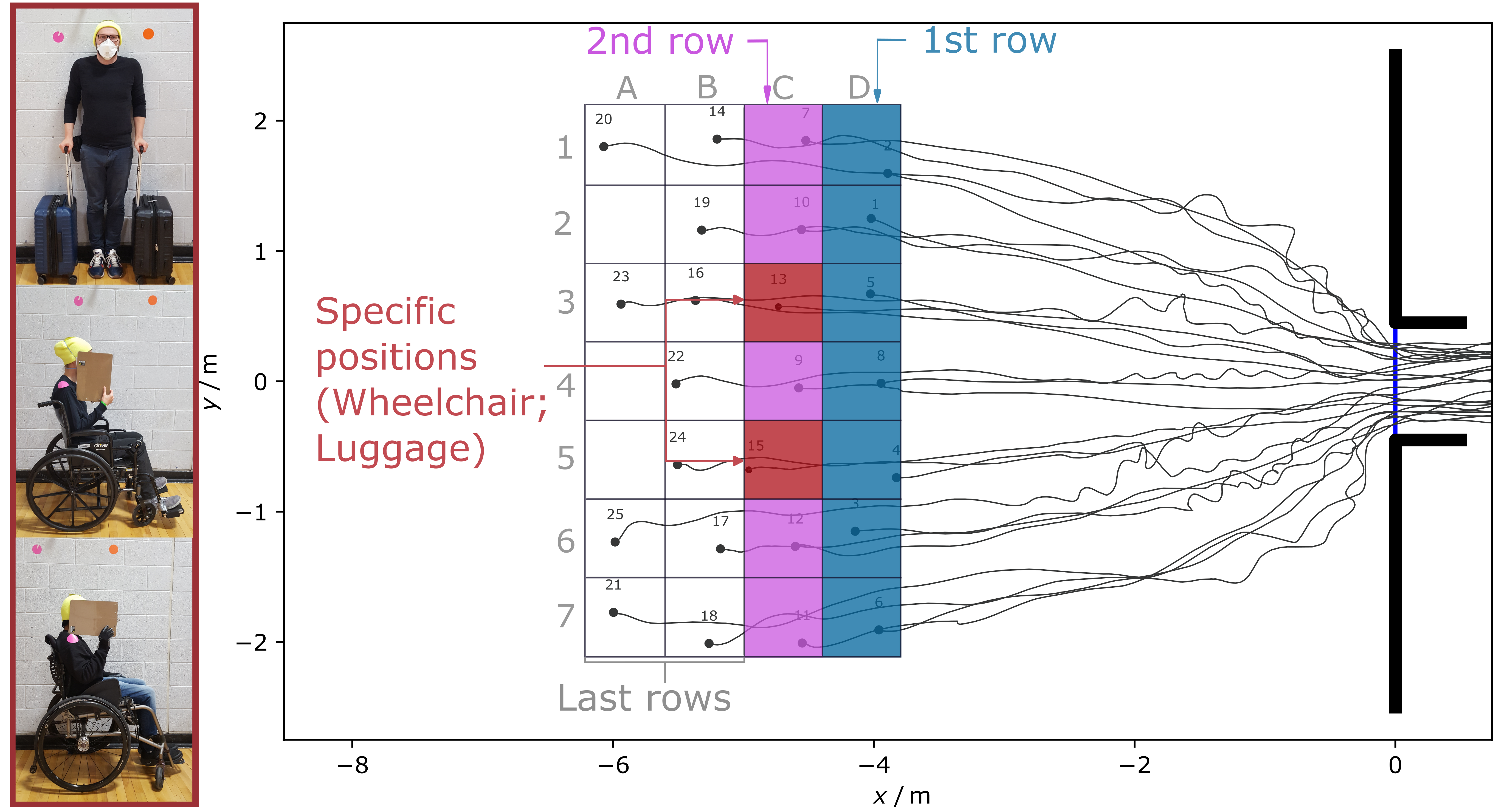} 
  \caption{Study configuration with predefined starting positions and participant trajectories.   Participants starting from specific positions (red colored area ($C3$, $C5$)) were either equipped with a manually operated wheelchair, carrying luggage, or were from the control group. }
  \label{fig:sketch} 
\end{figure}

\fref{fig:sketch} shows the study design. Participants had to stay in a regular marked area on the ground consisting of squared cells with a size of \SI{0.6}{\square\meter} starting \SI{4}{\meter} in front of the bottleneck. They moved through the bottleneck at a leisurely pace. We manipulated the mobility profile of two participants at the center of the crowd (either wheelchair users, pedestrians with two suitcases, or no manipulation). In the \textit{wheelchair condition}, two experimenters joined the study using manually operated wheel\-chairs. The other participants were not aware that these were neither naive participants nor everyday wheelchair users. Both had received adequate training in using a wheelchair.

In the \textit{luggage condition}, two spinner suitcases were randomly assigned to participants. The starting positions of participants using wheelchairs or carrying luggage were fixed to specific positions C3 and C5 (cf. \fref{fig:sketch}). 
The \textit{control condition} served as a reference point and the mobility profile of the two central participants was not modified. 

We differentiated between the following groups: (a) participants starting in the first row $D$, (b) participants starting at the defined starting positions $C3$ and $C5$ (central participants),  (c) participants without a wheelchair and without luggage, starting in the second row $C$ (but not $C3$ and $C5$), and (d) participants starting in the last two rows $A$ and $B$. Each trial was repeated for each condition about \SI{30}{} times. 

The passage of the bottleneck for each individual was measured at the blue line at $x=0$ (cf. \fref{fig:sketch}). Three metrics characterize the movement process upstream and inside the bottleneck: First, the time it takes for each participant to \textit{reach} $t_{\mathrm{reach}}$ the bottleneck, i.e., the time between the start of the trial and reaching the blue line. It describes the individual dynamics prior to reaching the bottleneck. Second, the time it takes each participant to \textit{pass through} $t_{\mathrm{pass}}$ the bottleneck, i.e., the time between the passage of a participant and his follower. This measure illustrates the dynamics within the bottleneck. Third, the time of the \textit{last participant} $t_{\mathrm{leave}}$ to pass the bottleneck in a trial, which characterizes the duration of the emptying process macroscopically for the entire crowd. 

\section{Results and discussion}
\label{sec:results}



\begin{figure}
  \centering
   \subfigure[\label{fig:perceivedvulnerability}]{\includegraphics[width=0.475\linewidth]{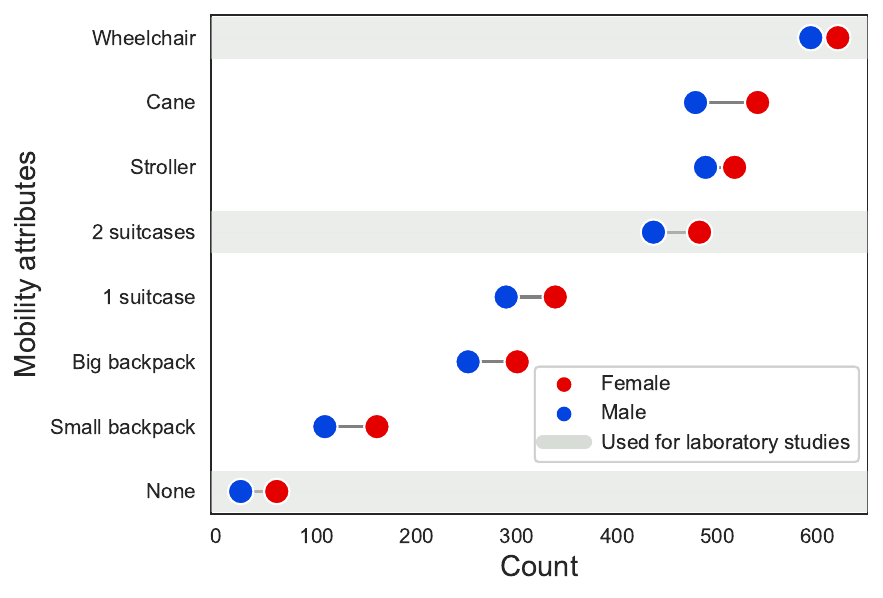}}
  \subfigure[\label{fig:timetoreach}]{\includegraphics[width=0.495\linewidth]{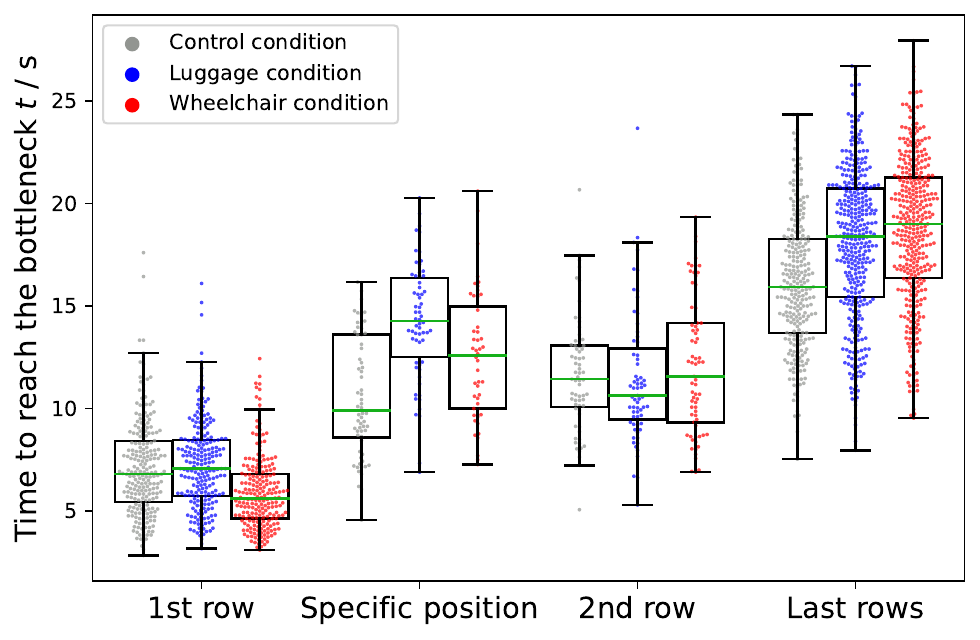}} \\
  \subfigure[\label{fig:timetopass}]{\includegraphics[width=0.495\linewidth]{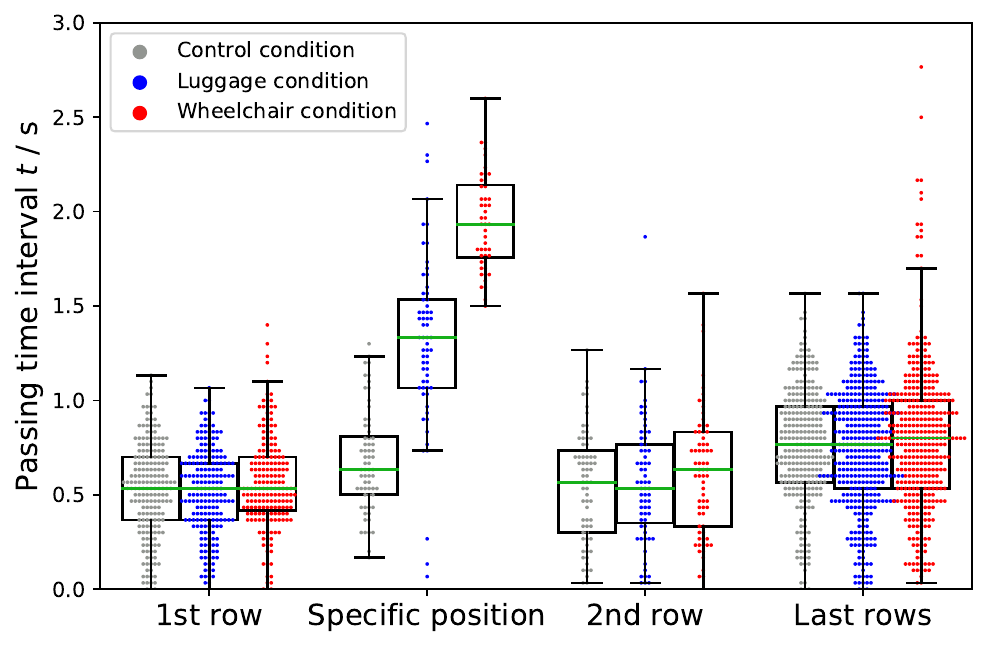}}
  \subfigure[\label{fig:timetoleave}]{\includegraphics[width=0.47\linewidth]{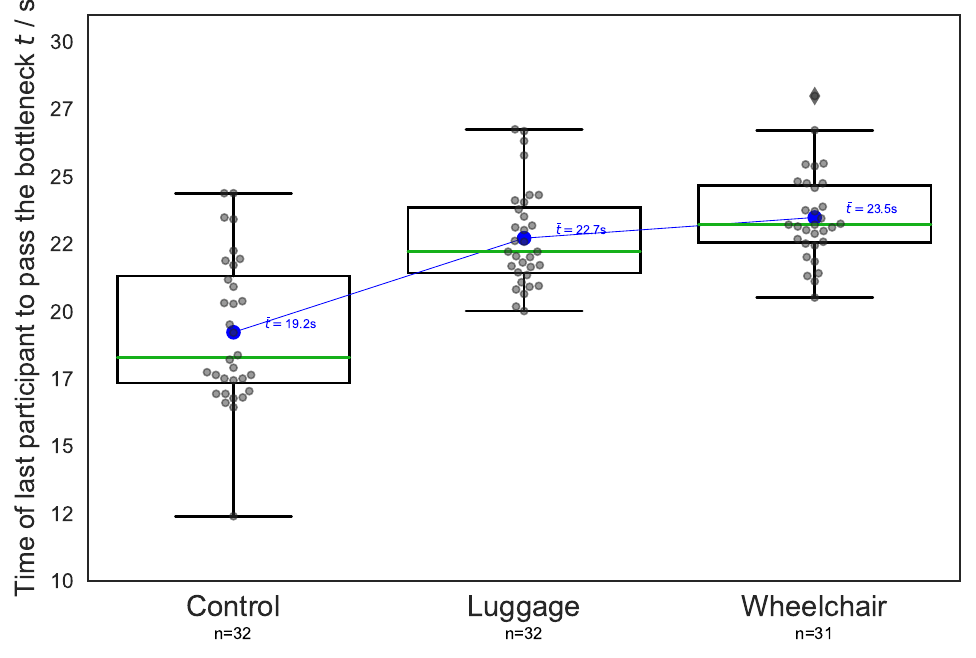}}
  \caption{(a) Perceived vulnerability as a function of mobility attributes. (b) Time to reach the bottleneck, (c) time to pass through the bottleneck, and (d) time of the last participant to pass the bottleneck as a function of start position and trial conditions. Green horizontal lines are the median, and blue points are the mean values. 
  }
  \label{fig:timetoreachandtimetopass} 
\end{figure}

\textbf{Time to reach the bottleneck}: Across conditions, participants in the first row were usually the first to reach the bottleneck. This is not particularly surprising, given that the first row was typically not affected by the movement of the two central participants. A smaller median ($\tilde{t}_{\mathrm{reach}}=\SI{5.6}{\second}$) was observed in the wheelchair condition compared to the control ($\tilde{t}_{\mathrm{reach}}=\SI{7.1}{\second}$) and luggage condition ($\tilde{t}_{\mathrm{reach}}=\SI{6.8}{\second}$). 
In the second row, participants in the control condition reached the bottleneck in a shorter time  ($\tilde{t}_{\mathrm{reach}} = \SI{9.9}{\second}$) than in the two other conditions. Wheelchair users reached the bottleneck faster ($\tilde{t}_{\mathrm{reach}} = $\SI{12.6}{\second}) than the luggage user ($\tilde{t}_{\mathrm{reach}} = \SI{14.3}{\second}$) which indicates that the experimental manipulation affected the movement of the central participants. It is somewhat striking that wheelchair users reached the bottleneck faster than luggage users. These differences perpetuate to the last rows, again with participants in the control condition reaching the bottleneck earlier than those in the other conditions. Overall, those starting from the last rows also needed the longest time to reach the bottleneck (\fref{fig:timetoreach}). 

\textbf{Time to pass through the bottleneck}: Generally, differences between starting positions were negligible (\fref{fig:timetopass}), suggesting that similar mobility characteristics lead to similar values. In the last rows, however, the median time to pass consistently exceeded those in the other rows by a small amount $\tilde{t}_{\mathrm{pass}} = \SI{0.53}{\second}$ in the first row vs. $\tilde{t}_{\mathrm{pass}} = \SI{0.77}{\second}$ in the last row). 
However, we observed that the experimental manipulations had strong effects on participants in the wheelchair and luggage condition. Carrying two suitcases doubled  ($\tilde{t}_{\mathrm{pass}} = \SI{1.3}{\second}$) and using a wheelchair even tripled ($\tilde{t}_{\mathrm{pass}} = \SI{1.9}{\second}$) the median time to \textit{pass} through the bottleneck compared to the control condition ($\tilde{t}_{\mathrm{pass}} = \SI{0.6}{\second}$). The outliers in the last row of the wheelchair condition likely were individuals whose time to pass through the bottleneck was affected by wheelchair or luggage users ahead of them.


\textbf{Last participant to leave the bottleneck}: In line with the previous observation, \fref{fig:timetoleave} shows that the last participants in the control condition ($\tilde{t}_{\mathrm{leave}} = \SI{18.3}{\second}$, $\bar{t}_{\mathrm{leave}} = \SI{19.2}{\second}$) were typically faster than those in the luggage ($\tilde{t}_{\mathrm{leave}} = \SI{22.2}{\second}$, $\bar{t}_{\mathrm{leave}} = \SI{22.7}{\second}$) and the wheelchair ($\tilde{t}_{\mathrm{leave}} = \SI{23.2}{\second}$, $\bar{t}_{\mathrm{leave}} = \SI{23.5}{\second}$) conditions. 

\section{Conclusion}
\label{sec:conclusion}

The results highlight the impact of individual mobility profiles on crowd dynamics. The presence of individuals with diverse characteristics adds complexity and interactivity to crowd movement. Specifically, when passing through the bottleneck, participants with luggage or in wheelchairs slowed down individually and affected those immediately behind them. However, those passing through the bottleneck \textit{before} them remained largely unaffected. It was most efficient to pass through the bottleneck without a wheelchair or luggage, and carrying luggage was less disruptive than using a wheelchair in such scenarios. 

Overall, our findings support previous work  (e.g., \cite{Geoerg.2022}, suggesting that increased heterogeneity changes crowd dynamics. This adds to the growing evidence challenging the status quo of existing approaches to estimate capacities and key performance values of crowd movement. The controlled approach to manipulate heterogeneity as well as the repeated trials in particular provide robust evidence that egress calculations that do not consider heterogeneous movement profiles likely generate overly optimistic results. 

However, more evidence is needed given the limitations of the present work, such as the relatively small sample and relying on simulated mobility impairments with participants who do not use wheelchairs in their daily lives. Consequently, the findings may not generalize to populations with varying mobility impairments and different types of mobility devices. This limitation underscores the need for future research to improve our understanding of crowd dynamics and visual confidence in augmented reality across a more diverse range of participants and conditions.

\begin{acknowledgement}
    The experiments were funded by the National Research Council Canada \textit{Ideation Fund - New Beginnings Initiative} (project number: NBR3-549).
    The authors would like to thank Carleton University, in particular, the Advanced Cognitive Engineering
    Laboratory, the Norm-Fenn gym staff, and the colleagues from the IAS-7 of the Forschungszentrum Jülich for their tremendous support. The authors also thank Sai Sumanth Neelamsetty, Oluchi Audu, Ben Jones, and the NRC Construction Research Center and Fire Safety team for their help and valuable support in making this data collection possible.
\end{acknowledgement}

\begin{ethics}
    All work conducted in this project was reviewed and approved by ethics committees. Approval was required from the NRC Research Ethics Board (\# 2021-106) as well as the ethics committees from Carleton University (\# 117454) and the University of Ottawa (\# H-04-22-8052). All researchers received training on research ethics before the study Ethical Conduct for Research Involving Humans (2018).
    Fundamentally, the study followed international guidelines on research involving human
    subjects, including procedures for informed consent, participant compensation, as well
    as the right to withdraw from the study at any point. 
\end{ethics}

\begin{contributions}
    Paul Geoerg: scientific content planning experiments, coordinating and technical realization experiments, data curation (camera, processing to trajectories, manual correction trajectories, combination of trajectories), writing (original draft, reviewing, editing) // 
    Ann Katrin Boomers:  technical realization of experiments, technical planning camera, data curation (camera, processing to trajectories, manual correction trajectories, combination of trajectories), scientific content planning experiments, writing (reviewing and editing original draft) // 
    Maxine Berthiaume: scientific content planning experiments, organisational realization experiments, data curation (participants data), writing (reviewing, editing) //
    Maik Boltes: supervision, technical planning camera, data curation camera, scientific content planning  experiments, writing (reviewing and editing original draft) //
    Max Kinateder: supervision, scientific content experiments, planning, coordinating, and technical responsibility experiments,  data curation (processing to trajectories, manual correction trajectories) writing (reviewing and editing original draft). 
\end{contributions}

\printbibliography


\end{document}